\documentclass[11pt,twoside]{article}

\usepackage{asp2006}
\usepackage{epsf}
\usepackage{psfig}
\usepackage{lscape}
\usepackage{graphicx}
\catcode`\@=11
\def\gsim{\ifmmode{\mathrel{\mathpalette\@versim>}}
    \else{$\mathrel{\mathpalette\@versim>}$}\fi}
\def\lsim{\ifmmode{\mathrel{\mathpalette\@versim<}}
    \else{$\mathrel{\mathpalette\@versim<}$}\fi}
\def\@versim#1#2{\lower 2.9truept \vbox{\baselineskip 0pt \lineskip
    0.5truept \ialign{$\m@th#1\hfil##\hfil$\crcr#2\crcr\sim\crcr}}}
\catcode`\@=12
\def\msun{\hbox{$M_\odot$}}
\def\lsun{\hbox{$L_\odot$}}
\def\yr-1{\hbox{${\rm yr}^{-1}$}}
\def\pn{\par\noindent}
\def\mpb{\medskip\pn$\bullet$\quad}
\def\pn{\par\noindent}

\begin{document}
\title{Witnessing Galaxy-SMBH Co-Evolution at Redshift $\sim 2$} 
\author{Alvio Renzini}   
\affil{INAF -- Osservatorio Astronomico di Padova, Italy}   
\author{Emanuele Daddi} 
\affil{Laboratoire AIM, CEA/DSM, France}
\begin{abstract} 
In a recent multiwavelength study of galaxies at $z\sim 2$ by Daddi et
al. (2007a,b) it is shown that galaxies with a Mid-IR excess most
likely harbor a Compton-thick AGN, thus bringing to $\sim 1/3$ the
fraction of $z\sim 2$ galaxies hosting an AGN. This finding opens a
number of intriguing issues concerning the concomitant growth of
galaxies and supermassive black holes, AGN feedback, and downsizing,
at the cosmic epoch of most intense star formation and nuclear
activity.

\end{abstract}


\section{Introduction}  
The current supermassive black hole (SMBH) and galaxy {\it
co-evolution} paradigma rests on three main arguments, namely: 

\mpb
Virtually all (massive) spheroids host a SMBH, 

\mpb The SMBH to stellar spheroid mass ratio $M_{\rm SMBH}/M_{\rm
spheroid}$ is $\sim 10^{-3}$, within a factor of two, known  as the
{\it Magorrian ratio} (Magorrian et al. 1998; Ferrarese et al. 2006).

\mpb 
We may need {\it AGN Feedback} to expell residual gas in spheroids,
switch-off star formation,  and start making the passively
evolving galaxies at $z\sim 2$ (e.g., Granato et al. 2001;
Bower et al. 2006). 

This paper presents recent evidence on SMBH/galaxy coevolution at
$z\sim 2$, i.e., at the peak of both galaxy growth and AGN activity,
as derived from a recent in depth study of a sample of $BzK$-selected,
$K_{\rm Vega}<22$ starforming galaxies in the GOODS Fields in the
range $1.4<z<2.5$ (Daddi et al. 2007a,b). At these redshifts these
galaxies account for $\sim 2/3$ of the total stellar mass, and for a
major fraction of the total star formation rate.

\section{Starforming and Mid-IR Excess Galaxies at $1.4\lsim z\lsim 2.5$}
Star formation rates (SFR) in high redshift galaxies can be estimated
in a number of ways, e.g. using the rest-frame UV, or the observed 24 $\mu$m, 
70 $\mu$m,  and sub-mm fluxes, or Radio and soft X-Ray data. Daddi et al. 
(2007a) have shown that for their sample of $\sim 1000$ $z\sim 2$ galaxies 
all these SFR indicators agree with each other within the errors, with one 
notable exception: the SFR estimated from Spitzer/MIPS 24  $\mu$m data.
This exception is illustrated by constructing the ratio:
\begin{equation}
R_{\rm SFR}={SFR_{\rm UV}({\rm uncorrected}) + SFR_{24\mu{\rm m}}\over
  SFR_{\rm UV} ({\rm dust-corrected})},
\end{equation}
 where $SFR_{\rm UV}({\rm uncorrected})$ is the SFR from the rest
frame 1500 \AA \ flux {\it before} applying the extinction correction,
$SFR_{\rm UV}({\rm dust-corrected})$ is the SFR as derived from the
rest frame 1500 \AA \ flux {\it after} applying the extinction
correction (estimated from the UV slope), and $ SFR_{24\mu{\rm m}}$ is the
SFR as derived from the Spitzer/MIPS 24 $\mu$m data, i.e., from the
rest-frame $\sim 8\;\mu$m flux plus a model SED (e.g., Chary \& Elbaz,
2001). Thus, this is meant to represent the ratio of the un-extincted
SFR plus the extincted SFR, over the SFR corrected for extinction.
Fig. 1 shows this ratio as a function of redshift for the sample
galaxies.

\begin{figure}[!t]
\centering
\includegraphics[width = 8cm, angle = -90]{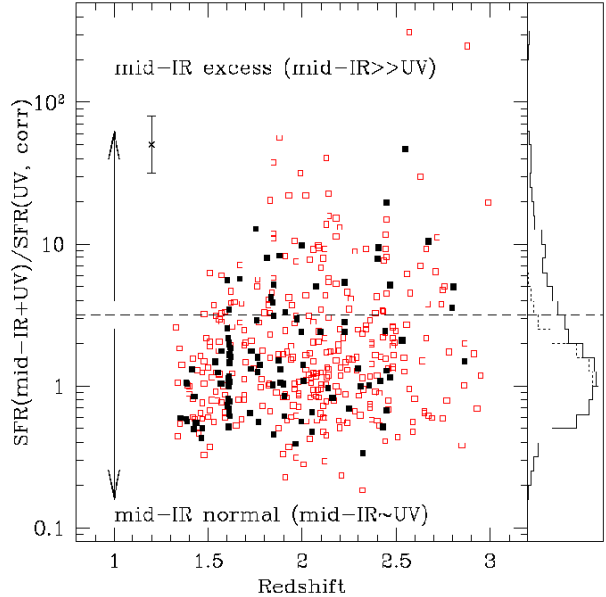}
\caption{The identification of the Mid-IR excess galaxies over
the GOODS-South field. The ratio defined in Eq. (1) is plotted vs redshift.
Open and closed symbols refer to photometric and
spectroscopic redshifts, respectively. (from Daddi
et al. 2007b).}
\label{fig:1}       
\end{figure}
If both SFR estimates were unbiased, one would expect the distribution
of this ratio to be like a Gaussian, peaking at $R_{\rm SFR}=1$, and
be symmetric: the result of random errors in either estimates. Fig. 1
shows that the distribution does indeed peak at $R_{\rm SFR}=1$, but
the distribution is skewed towards large values of $R_{\rm SFR}$. A
Gaussian fitting the $R_{\rm SFR}\le 1$ part of the distribution is
also shown in Fig. 1, allowing to quantify in $\sim 25\%$ of the total
the fraction of high $R_{\rm SFR}$ galaxies, in excess of the Gaussian
distribution. Thus, Daddi et al. isolate a population {\it Mid-IR
Excess} galaxies, i.e., those for which the rest-frame 8 $\mu$m flux 
overestimates the SFR, or, equivalently, the dust 24 $\mu$m emission
is in excess of what is expected from the extinction as estimated from
the UV slope.
\section{Compton-Thick AGNs are Widespread at $z\sim 2$}
In principle, the excess SFR estimated from the 24 $\mu$m flux may
signal some inadequacy of the semi-empirical relation used to
translate the rest frame $\sim 8\;\mu$m flux into a
SFR. Alternatively, the Mid-IR excess may signal that dust is being
heated also by an energy source other than star formation, i.e.,
nuclear activity. 

To explore this latter option, Daddi et al. (2007b) have appealed to
Chandra X-ray data publicly available over the CDFS/GOODS field, proceeding
to stack the X-ray images, separately for the {\it normal} ($R_{\rm
SFR}<3$) and the Mid-IR excess ($R_{\rm SFR}>3$) galaxies, not
including those galaxies which are individually detected in
X-Rays. The result is shown in Fig. 2. Clearly, in the stacked images
the normal galaxies are well detected in the soft band, but just
barely in the hard band. The Mid-IR excess galaxies, instead, show
strong emission also in the hard band, which is indeed interpreted as
evidence for them hosting an X-ray emitting AGN.
\begin{figure}[!t]
\centering
\includegraphics[width = 8cm, angle = -90]{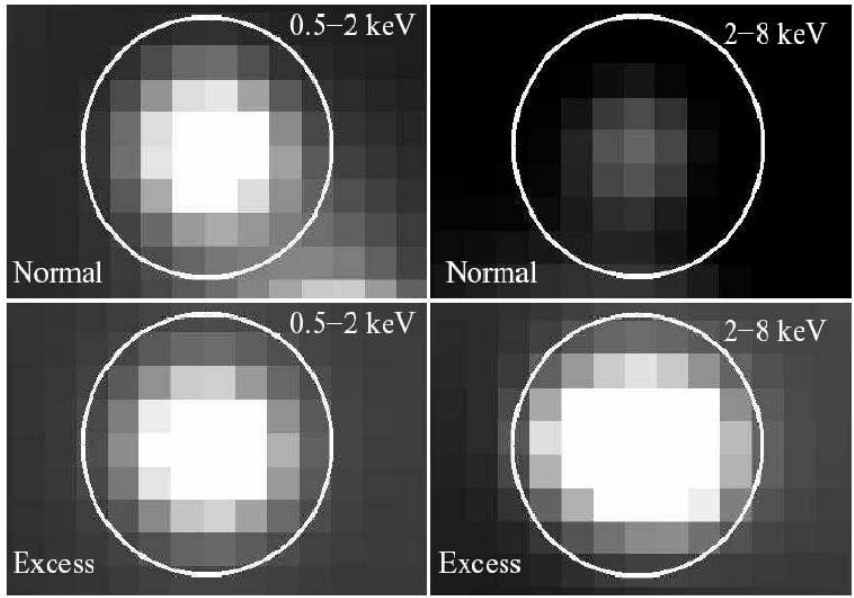}
\caption{The Chandra stacked images in the soft (0.5--2 keV) and hard
(2-8 keV) X-ray bands, separately for the normal and the Mid-IR excess
galaxies (from Daddi et al. 2007b).}
\label{fig:2}       
\end{figure}
From the fit of their X-ray spactrum to those of model AGN spectra
with different column densities of absorming material one
infers that the AGN X-ray flux is heavily absorbed in the soft
bands, as illustrated in Fig. 3, with log $N_{\rm H}\gsim 24.5$,
indicating that the majority of such AGNs are Compton-thick.
The model X-ray spectra by Gilli, Comastri \& Hasinger (2007) 
used for the fit are shown in Fig. 4 (left panel).
\begin{figure}[!t]
\centering
\includegraphics[width = 6 cm,angle = -90]{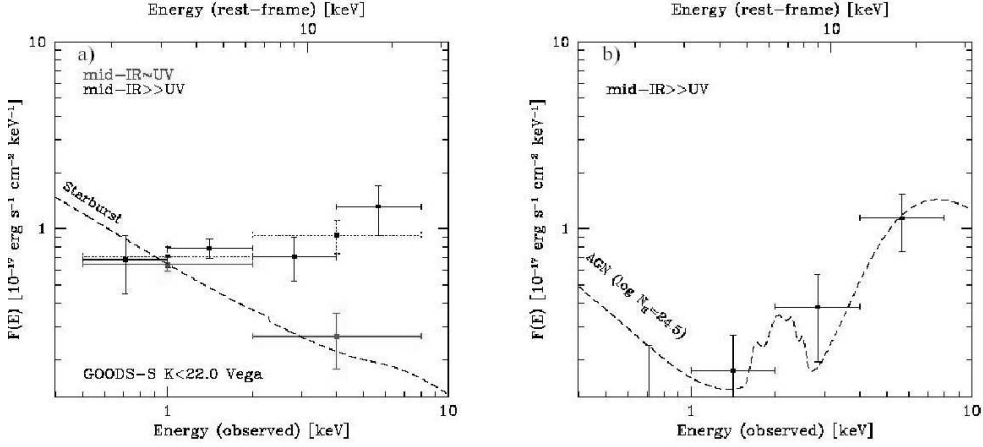}
\caption{Left panel: The Chandra X-ray spectrum of stacked normal
(red) and the Mid-IR excess (black) galaxies. The dashed line refers
to the spectrum of a pure starburst galaxy. Right panel: the X-ray
spectrum of the stacked Mid-IR excess galaxies, after subtraction of
the soft component due to star formation. Also show is the model X-ray
spectrum of an heavily absorbed AGN from Gilli, Comastri \& Hasinger
(2007), with a column density log $N_{\rm H}=24.5$, qualifying the
typical Mid-IR excess galaxy as Compton-thick.
(from Daddi et al. 2007b).}
\label{fig:3}       
\end{figure}

Thus, Daddi et al. infer that at $z\sim 2$ some 25\% of $K_{\rm
Vega}<22$ galaxies host a Compton-thick AGN, and over $\sim 35\%$
hosts an AGN, when including also those galaxies that are individually
detected in X-rays (about one half in number compared to the Mid-IR
excess galaxies to the same $K_{\rm Vega}<22$ limit, Daddi et
al. 2007b). Clearly, nuclear activity was vastly more widespread at
$z\sim 2$ compared to the local universe. For a similar result see
also Fiore et al. (2008).

\section{The concomitant growth of the stellar mass and the BH mass}

The $1.4<z<2.5$ redshift range, or the cosmic time from $\sim
2.5$ to $\sim 4.5$ Gyr since the Big Bang, marks the epoch of most
intense star formation and nuclear activity. Hence, it must be the
epoch when {\it co-evolution} of galaxies and SMBHs should be most
prominent and, as such, most easily recognizable.

This is quantitatively explored in Daddi et al. (2007b). First, by
estimating the average SFR over the whole sample of $K<22$ galaxies at
$1.4<z<2.5$, for which one derives $<\! SFR\!>=70\;\msun\yr-1$.  The
SMBH mass-growth rate can be estimated from the unabsorbed X-Ray
luminosity and a canonical $\sim 0.1$ efficiency for the accreted
mass-energy conversion. Over the whole $K<22$ sample (i.e., including
all galaxies) the average unobscured X-ray luminosity is
$10^{11}-10^{12}\lsun$, hence $<\! M_{\rm SMBH}/dt\!> =
0.025-0.25\;\msun\yr-1$, and therefore:
\begin{equation}
{<\! dM_{\rm SMBH}/dt\!>\over <\! SFR\!>} = (0.35-3.5)\times 10^{-3},
\end{equation}
so tantalizingly bracketing the {\it Magorrian ratio} ($\sim
10^{-3}$).  Indeed, {\bf are we seeing the Magorrian ratio being
established in this sample of $z\sim 2$ galaxies?}

\begin{figure}[!t]
\centering
\includegraphics[width=6cm, angle = 0]{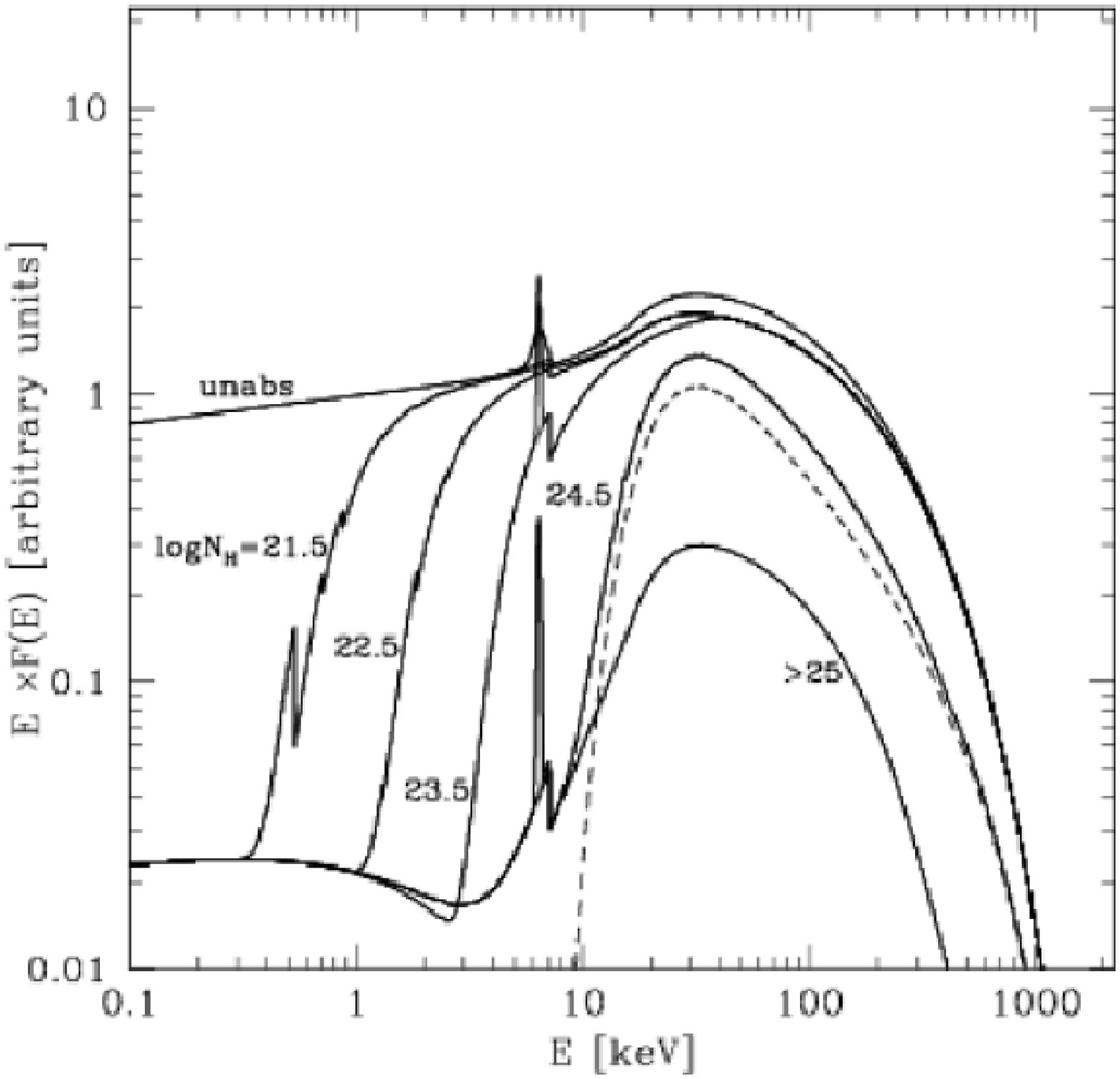}
\hspace{\fill}
\includegraphics[width=6cm, angle = 0]{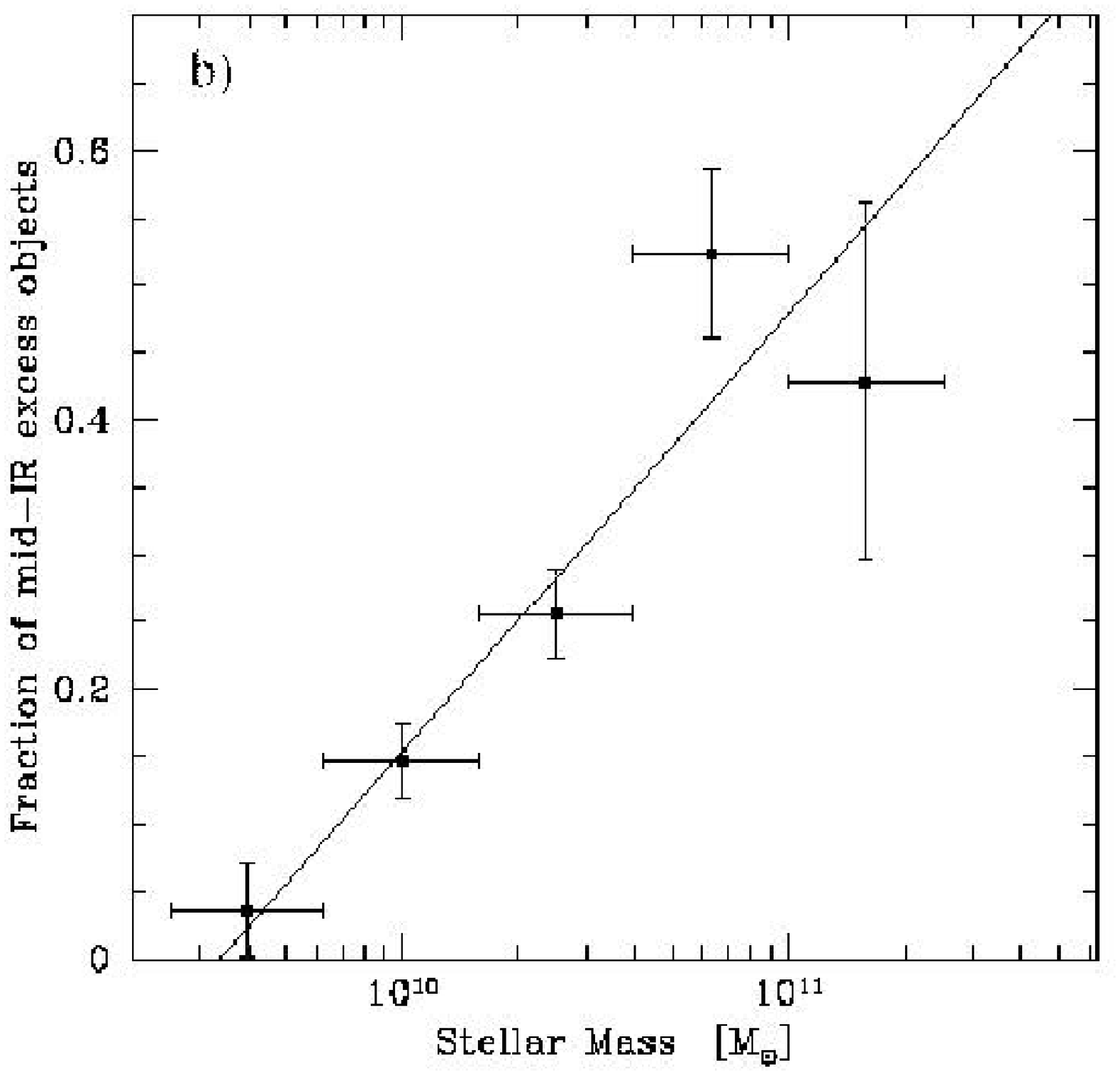}
\vskip 2mm
\caption{ Left panel: the model AGN spectra for different column
densities of the absorming material. The assumed AGN unabsorbed
spectrum is labelled ``unabs'' (From Gilli, Comastri \& Hasinger
2007). Right panel: the fraction of Mid-IR excess galaxies as a
function of stellar mass, for the $K<22$ sample of $z\sim 2$
galaxies.}
\label{fig:4}       
\end{figure}

\section{AGN Feedback}

A necessary condition for AGN feedback is having an AGN. With $\gsim
35\%$ of all $K<22$ galaxies hosting an AGN it looks that this
condition is quite well satisfied at $z\sim 2$. Feedback arises
because energy (momentum) is transfered from the AGN to the ISM of the
host galaxy, heating it and eventually expelling it from the galaxy.
This energy transfer may be mediated by relativistic jets, or by hard
AGN radiation. We don't know whether relativistic jets are buried
inside these galaxies, but we have direct evidence from the X-rays
that hard radiation is certainly at work.  In fact, the second
condition --energy transfer from the AGN to the ISM-- is also
satisfied: being Compton-thick, a major fraction of the X-ray
luminosity of the AGNs in the Mid-IR excess galaxies is clearly dumped to the
ISM.  The unobscured, hard X-Ray luminosity (at $>6$ keV in the
rest-frame) of these Compton-thick AGNs is typically $\sim 10^{43}$
erg/s, a power which is being absorbed/deposited in the ISM. Formally,
this power would be sufficient to eject all the gas from a galaxy in
just a few million years. However, most of the energy deposited in the
ISM must be degraded locally, perhaps even in the immediate vicinity of the
AGN.  Indeed, the mere Mid-IR excess testifies that the hard X-ray
photons Compton-heat electrons in the ISM, which share their energy
with gas and dust, and eventually dust radiates most of this energy
away in the Mid- and Far-IR. From the point of view of feedback, this
is a pure loss. Nonetheless, if even a relatively small fraction of 
the absorbed hard X-ray luminosity (say, a few percent) goes to
increasingly Compton-heat the gas, then SF could be quickly switched
off, gas expelled from the galaxy, and the galaxy itself turned into
an early-type, passively evolving galaxy.

These, considerations prompt a second, tantalizing question: {\bf with these
Compton-thick AGNs, are we seeing the AGN Feedback in action?}

\section{Downsizing}
Fig. 4 (right panel) shows the fraction of Mid-IR excess galaxies
(those with $R_{\rm SFR}>3$) as a function of their stellar mass. This
fraction increases with mass, reaching $\sim 50\%$ at $M\sim
10^{11}\;\msun$. Thus, Compton-thick AGN activity, and the likely
feedback) are even more widespread among the most massive star-forming
galaxies. It is therefore likely that AGN feedback will more promptly
succed in quenching star formation in the most massive galaxies,
whereas it may take longer to activate a powerful Compton-thick AGN in
progressively lower-mass galaxies, with star formation persisting in
them for a longer time.

It is now observationally well established that the most massive
galaxies are the first to turn passive, starting at $z\gsim 2$, while
less massive galaxies turn passive at lower and lower redshifts the
lower their mass (e.g., Kodama et al. 2004; Thomas et al. 2005;
Cimatti, Daddi \& Renzini 2006), a manifestation of {\it Galaxy
Downsizing}. Thus, we meet here the third tantalizing question
promoted by this analysis of $z\sim 2$ galaxies: {\bf Are we seeing 
``Galaxy Downsizing'¡É in progress?} 

\section{Conclusions}
The systematic study of Daddi et al. (2007a,b) of a complete sample of 
$1.4<z<2.5$ galaxies has revealed a widespread Compton-thick AGN activity
among these galaxies, and has prompted three intriguing questions, namely:

\mpb
Are we seeing the Magorrian ratio being
established in $z\sim 2$ galaxies?
\mpb
Among these galaxies are we seeing the AGN Feedback in action?
\mpb
Are we seeing ``Galaxy Downsizing'¡É in prograss?

\pn 
We cannot yet give a secure, positive answer to these questions,
as much remains to be explored on the structure and internal workings
of these galaxies. However, with cosmic SFR and AGN activity both
peaking at $z\sim 2$, if not at this redshift (i.e., in the ``boom
years'') when/where would we better witness galaxy-SMBH co-evolution?



\end{document}